\documentclass[twocolumn,prl,preprintnumbers,amsmath,amssymb,groupedaddresst]{revtex4}
\usepackage{graphicx}
\usepackage{color}
\begin{document}

\preprint{RIKEN-MP-16}

\title{Anomaly-induced charges in nucleons}

\author{Minoru  {\sc Eto}\footnote{At Yamagata University since April 2011.}}\author{Koji {\sc Hashimoto}}\author{Hideaki {\sc Iida}}\author{Takaaki {\sc Ishii}\footnote{At University of Cambridge since April 2011.}}
\author{Yu {\sc Maezawa}}
\email[]{meto, koji, hiida, ishiitk(at)riken.jp; maezawa(at)ribf.riken.jp}
\affiliation{
{\it Mathematical Physics Lab., RIKEN Nishina Center, Saitama 351-0198, Japan }}

\begin{abstract}
We show a novel charge structure of baryons in electromagnetic 
field due to the chiral anomaly. 
A key connection is to treat baryons as solitons of mesons. 
We use Skyrmions to calculate the charge distributions in a single nucleon 
and find an additional charge. 
We also perform calculations of charge distribution for classical multi-baryons with $B=2, 3, \cdots,8 \ {\rm and} \ 17$; 
they show amusing charge distributions.  
\end{abstract}

\maketitle


Recent advance in observations and experiments explores 
new effects of strong electromagnetic fields 
on fundamental particles. Since matter consists of baryons,
electromagnetic properties of protons and neutrons are of most importance.
Under the strong magnetic fields such as in neutron stars, supernovae
and heavy ion collisions, tiny quantum effect of quantum chromo-dynamics may
lead to an unveiled and significant consequence.

In this letter, 
we investigate baryons
under external electromagnetic fields. 
For describing the baryons, we use the Skyrme model \cite{Skyrme1961} 
with Wess-Zumino-Witten (WZW) term \cite{Wess1971,Witten1983} 
including electromagnetism. 
The consequence is amazing: 
Nucleons in the external electromagnetic fields have anomalous charge distribution due to the chiral anomaly. 
Nonzero net charge, which is generally non-integer, is induced {\it even for neutrons}. 
Correspondingly, we will show that the Gell-Mann-Nishijima formula, $Q=I_3+N_B/2$ ($Q$: electric charge, $I_3$: 
the third component of isospin, $ N_B$: baryon number), has an additional term due to the quantum anomaly.  

\begin{figure}[t]
\includegraphics[width=0.28\textwidth]{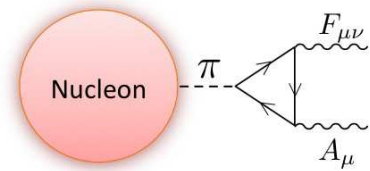}
\caption{A schematic figure for electric charge generation of a nucleon. 
In electromagnetic backgrounds, i.e., $F_{\mu\nu}\neq 0$, 
 the quark-loop diagram generates an additional 
coupling to the gauge fields $A_\mu$. }
\label{chargegeneration}
\end{figure}

Figure \ref{chargegeneration} shows a schematic description of the phenomenon. 
Due to the anomalous interaction with a quark loop through the WZW term, 
nucleons (= Skyrmions) have an additional interaction to the electromagnetic field $A_\mu$.  
Under the external electromagnetic fields, the anomalous coupling induces the additional electric charge. 

The phenomenon is not counter-intuitive. 
For example, in the Witten effect \cite{Witten1979b,Wilczek1987}, monopoles carry an electric charge 
under a nontrivial axion field configuration $\theta(\vec x,t)$ 
via an anomalous CP-odd term, $\theta \tilde F^{\mu\nu} F_{\mu\nu}$. 
The chiral magnetic effect (CME) \cite{Kharzeev2008,Fukushima2008,Eto2010,Voloshin2009b} 
is also a similar effect in heavy-ion collisions. 

\vspace{2mm}
{\noindent \it Meson effective action, Skyrmions and anomaly.} ---
We adopt the Skyrme model for a concrete illustration in this letter. 
The essential idea of the Skyrme model is to unify baryons and mesons:
 baryons are described as topological solitons of mesons.  
This model, known to reproduce experimental data of nucleons within 30\% accuracy,
is suitable for our purpose. This is because 
we concentrate on the anomalous contribution to baryons, which is   
described by the coupling between mesons and photons shown in Fig.~\ref{chargegeneration}. 
Any baryons wearing mesonic clouds will follow our 
mechanism of anomalous charge generation. 

The action of two-flavor Skyrme model \cite{Skyrme1961,Adkins1983,Adkins1984} coupled with 
electromagnetic field is 
\begin{align}
&S=\int d^4 x \left({\cal L}_{\rm kin}+{\cal L}_{\rm mass} -\frac{1}{4} F_{\mu\nu}F^{\mu\nu}\right)\label{action},\\
&{\cal L}_{\rm kin}=-\frac{F_\pi^2}{16}{\rm tr} (\hat R_\mu \hat R^\mu) +\frac{1}{32 e_s^2}
{\rm tr}([\hat R_\mu, \hat R_\nu][\hat R^\mu, \hat R^\nu]), \nonumber\\
&{\cal L}_{\rm mass}=\frac{F_\pi^2}{16}m_\pi^2{\rm tr} (U+U^\dagger -2), \hspace{5mm}
\hat R_\mu=D_\mu U U^\dagger, \nonumber 
\end{align}
where $m_\pi$ and $F_\pi$ are the pion mass and the pion decay constant, respectively. 
$e_s$ is a dimensionless constant and 
$D_\mu U\equiv \partial_\mu U+ ie A_\mu [q,U]$ with $q\equiv {\rm diag}(2/3,-1/3)$.  
The pseudo-scalar field $U$ is an SU(2) matrix which 
transforms as $U\rightarrow G_L U G_R^\dagger$ with $G_L\in {\rm SU(2)}_L$ and $G_R\in {\rm SU(2)}_R$.  
In the following, we use dimensionless variables: 
$r\rightarrow r/(e_s F_\pi)$ and $m_\pi\rightarrow (e_s F_\pi) m_\pi$.   
 In the Skyrme model, a general hedgehog-type ansatz in the absence of the electromagnetic 
 background is written as 
\begin{align}
  U&=GU_0G^\dagger
 =G\exp(if(r)\hat {\boldsymbol x}\cdot {\boldsymbol \tau})G^\dagger,\label{genehedgehog} \\
 G&=a_0+i{\boldsymbol a}\cdot {\boldsymbol \tau} \in {\rm SU(2)}_{L+R}, \ \ (a_0^2+{\boldsymbol a}^2=1) ,\label{geneSU2}
 \end{align}
 where $\hat{\boldsymbol x}\equiv {\boldsymbol x}/|{\boldsymbol x}|$,   
 ${\boldsymbol \tau}$ are Pauli matrices, and $(a_0,{\boldsymbol a})$ are moduli parameters 
 spanning ${\rm SU(2)}_{L+R} \simeq S^3$. 
 We treat electromagnetic effects as a perturbation in terms of $e$. 
The equation of motion gives
\begin{align}
\left(\frac{r^2}{4}+2\sin^2 f\right)f^{\prime\prime}+\frac{r}{2}f^{\prime}
+\sin(2f)\left( f^{'2}-\frac{1}{4}-\frac{\sin^2 f}{r^2}\right) \nonumber \\
-\frac{m_\pi^2 r^2}{4}\sin f =0. \nonumber
\end{align} 
Solving this under the boundary conditions, 
$f(0)=\pi$ and $f(r\rightarrow \infty)=0$, 
one obtains a solution with baryon number $B=1$. 
The solution is a topological soliton, called Skyrmion. 

We focus on the coupling between mesons and photons   
in the WZW term. 
In the two-flavor case, this can be given by \cite{Son2008,Witten1983,Wess1971}
\begin{align}
S_{\rm WZW}[A_\mu]&=-\int d^4x \ A_\mu \left( \frac{eN_c}{6}j^\mu_B
+\frac{1}{2} j_{{\rm anm}}^\mu\right)\label{WZWaction},\\
j^\mu_B&=\frac{1}{24\pi^2}\epsilon^{\mu\nu\rho\sigma}
{\rm tr}[R_\nu R_\rho R_\sigma]\label{baryoncurrent},\\
j^\mu_{{\rm anm}}&=-\frac{ie^2N_c}{96\pi^2}\epsilon^{\mu\nu\rho\sigma}F_{\nu\rho}{\rm tr}
[\tau_3(L_\sigma+R_\sigma)]\label{anmcurrent},
\end{align}
where $j^\mu_B$ is a baryon current giving an integer baryon number, 
$L_\mu=U^\dagger \partial_\mu U$, 
$R_\mu=\partial_\mu U U^\dagger$, 
and $\epsilon^{0123}=-1$ in this letter. 

In the presence of background electromagnetic fields, 
not only the first term but also the second term in Eq.~(\ref{WZWaction}) is important.   
The electric charge $Q$ with the contribution from anomaly  
($N_c=3$) is written as 
\begin{align}
Q=I_3+\frac{N_B}{2}+\frac{Q_{{\rm anm}}}{2},\label{modNG}
\end{align}
where $N_B=\int d^3x j^0_B$ and $Q_{{\rm anm}}=\int d^3x j^0_{{\rm anm}}$.
Thus, the Gell-Mann-Nishijima formula is corrected 
under background electromagnetic fields. 
Substituting Eq.~(\ref{genehedgehog})  
into Eq.~(\ref{anmcurrent}), 
we obtain 
\begin{align}
j^\mu_{{\rm anm}}=&
\dfrac{ie^2N_c}{96\pi^2}\epsilon^{\mu\nu\rho\sigma}F_{\nu\rho} P_\sigma, \label{chargedistcls}\\
P_\mu=&4i\left[(\partial_\mu f)\hat x^{\rm rot}_3+\frac{\sin(2f)}{2}\partial_\mu(\hat x^{\rm rot}_3)\right],\\
\hat x_3^{\rm rot}=&(a_0^2+a_3^2-a_1^2-a_2^2)\hat x_3
                                 \nonumber \\
                                 &+2(a_1a_3+a_0a_2)\hat x_1
                                 +2(a_2a_3-a_0a_1)\hat x_2.\label{genex}
\end{align}
This is a classical anomalous current for the general hedgehog solutions. 

\vspace{2mm}\noindent{\it Induced charges from quantized Skyrmion.} 
--- To obtain physical values of the anomalous charge depending on the baryon 
states, we need to quantize the Skyrmion. 
By solving a quantum mechanics of the $S^3$ 
moduli parameters on the Skyrmion,  
quantum states of a nucleon with spin quantized along $x^3$ 
are given by $\psi_{p \uparrow} = (a_1+ia_2)/\pi$, etc.\cite{Adkins1983}. 
We evaluate matrix elements of the anomalous current 
\footnote{The off-diagonal matrix elements contribute 
to the transition among the states. The matrix elements between the states with different $S_3$ survive,  
while those with different $I_3$ vanish.}, 
\begin{align}
\langle  j^\mu_{{\rm anm}} \rangle_{I_3, S_3} \equiv \int d\Omega_3 
 \psi^*_{I_3, S_3} \ j_{{\rm anm}}^{\mu}(a_0, {\boldsymbol a}) \ \psi_{I_3, S_3}, \label{matrixelement}
\end{align}
where $\int d\Omega_3$ denotes the integration over the $S^3$, 
and $\psi_{I_3, S_3}$ is the baryon states labeled by the third components of 
isospin and spin. 
The matrix elements are calculated as 
\begin{align}
\langle j_{{\rm anm}}^\mu\rangle_{I_3, S_3}=&
\ \dfrac{ie^2N_c}{96\pi^2}\epsilon^{\mu\nu\rho\sigma}F_{\nu\rho}\langle P_\sigma\rangle_{I_3,S_3},\nonumber\\
\langle P_0\rangle_{I_3,S_3}=&\ 0\nonumber,\\
\langle P_a\rangle^{a=1,2}_{I_3,S_3}=&-\frac{16i}{3}I_3 S_3  \left( f^\prime -\frac{\sin(2f)}{2r}\right) \hat x_a\hat x_3, 
\nonumber \\
\langle P_3\rangle_{I_3,S_3}=&-\frac{16i}{3}I_3 S_3 \left[ \left( f^\prime -\frac{\sin(2f)}{2r}\right) \hat x_3^2+\frac{\sin(2f)}{2r}\right]. \nonumber
\end{align}

\begin{figure}[t]
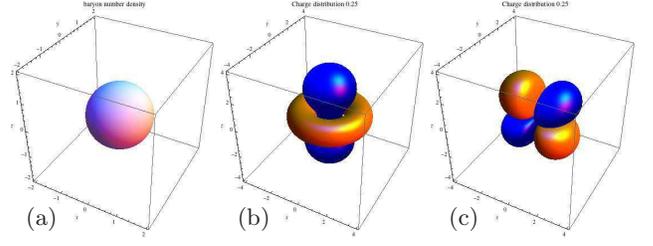

\includegraphics[width=0.15\textwidth]{./b1_01.eps2}
\includegraphics[width=0.15\textwidth]{./b1_02.eps2}
\includegraphics[width=0.15\textwidth]{./b1_05.eps2}
\put(-230,5){(a)}
\put(-150,5){(b)}
\put(-70,5){(c)}
\caption{
The constant-height surfaces of (a) density distribution of baryon number,  (b) electric charge 
under magnetic field along the 3rd-axis, and (c) electric charge under magnetic field 
along the 1st-axis. We used $f(r)$ for $m_\pi=0$. 
In (b) and (c), colors stand for positive and negative charge distributions. 
}
\label{B1chargedist}
\end{figure}

\begin{figure*}[t]
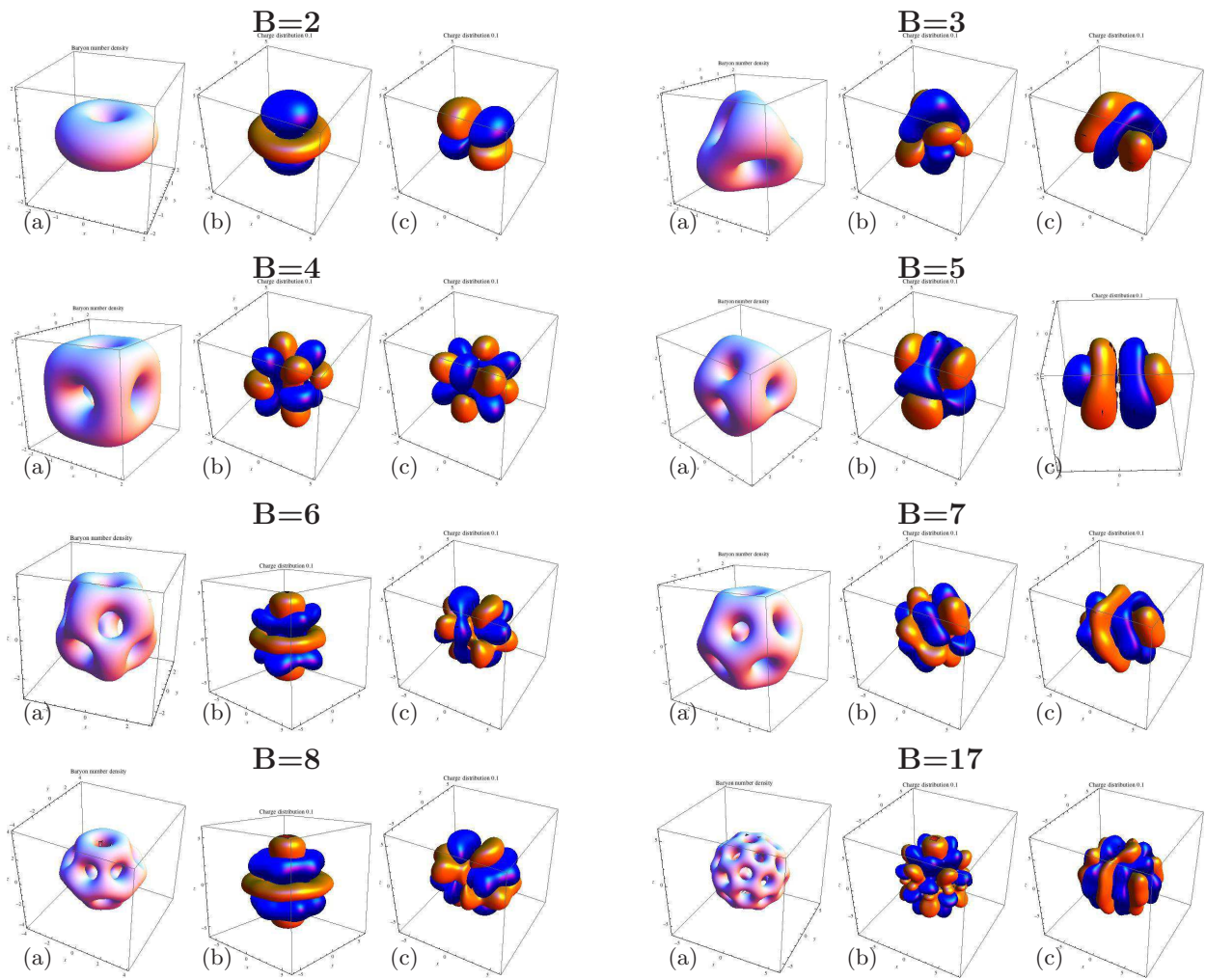

\begin{minipage}{.49\linewidth}
\vspace{0.18cm}
\includegraphics[width=0.28\textwidth]{./b2_01.eps2}
\includegraphics[width=0.28\textwidth]{./b2_02.eps2}
\includegraphics[width=0.28\textwidth]{./b2_05.eps2}
\put(-120,82){\large\bf B=2}
\put(-210,5){(a)}
\put(-140,5){(b)}
\put(-66,5){(c)}
\end{minipage}
\begin{minipage}{.49\linewidth}
\vspace{0.18cm}
\includegraphics[width=0.28\textwidth]{./b3_01.eps2}
\includegraphics[width=0.28\textwidth]{./b3_02.eps2}
\includegraphics[width=0.28\textwidth]{./b3_05.eps2}
\put(-120,82){\large\bf B=3}
\put(-210,5){(a)}
\put(-140,5){(b)}
\put(-66,5){(c)}
\end{minipage}
\begin{minipage}{.49\linewidth}
\vspace{0.18cm}
\includegraphics[width=0.28\textwidth]{./b4_01.eps2}
\includegraphics[width=0.28\textwidth]{./b4_02.eps2}
\includegraphics[width=0.28\textwidth]{./b4_05.eps2}
\put(-120,82){\large\bf B=4}
\put(-210,5){(a)}
\put(-140,5){(b)}
\put(-66,5){(c)}
\end{minipage}
\begin{minipage}{.49\linewidth}
\vspace{0.18cm}
\includegraphics[width=0.28\textwidth]{./b5_01.eps2}
\includegraphics[width=0.28\textwidth]{./b5_02.eps2}
\includegraphics[width=0.28\textwidth]{./b5_05.eps2}
\put(-120,82){\large\bf B=5}
\put(-210,5){(a)}
\put(-140,5){(b)}
\put(-66,5){(c)}
\end{minipage}
\begin{minipage}{.49\linewidth}
\vspace{0.18cm}
\includegraphics[width=0.28\textwidth]{./b6_01.eps2}
\includegraphics[width=0.28\textwidth]{./b6_02.eps2}
\includegraphics[width=0.28\textwidth]{./b6_05.eps2}
\put(-120,82){\large\bf B=6}
\put(-210,5){(a)}
\put(-140,5){(b)}
\put(-66,5){(c)}
\end{minipage}
\begin{minipage}{.49\linewidth}
\vspace{0.18cm}
\includegraphics[width=0.28\textwidth]{./b7_01.eps2}
\includegraphics[width=0.28\textwidth]{./b7_02.eps2}
\includegraphics[width=0.28\textwidth]{./b7_05.eps2}
\put(-120,82){\large\bf B=7}
\put(-210,5){(a)}
\put(-140,5){(b)}
\put(-66,5){(c)}
\end{minipage}
\begin{minipage}{.49\linewidth}
\vspace{0.18cm}
\includegraphics[width=0.28\textwidth]{./b8_01.eps2}
\includegraphics[width=0.28\textwidth]{./b8_02.eps2}
\includegraphics[width=0.28\textwidth]{./b8_05.eps2}
\put(-120,82){\large\bf B=8}
\put(-210,5){(a)}
\put(-140,5){(b)}
\put(-66,5){(c)}
\end{minipage}
\begin{minipage}{.49\linewidth}
\vspace{0.18cm}
\includegraphics[width=0.28\textwidth]{./b17_01.eps2}
\includegraphics[width=0.28\textwidth]{./b17_02.eps2}
\includegraphics[width=0.28\textwidth]{./b17_05.eps2}
\put(-120,82){\large\bf B=17}
\put(-210,5){(a)}
\put(-140,5){(b)}
\put(-66,5){(c)}
\end{minipage}
\caption{The corresponding plots of Fig.~2 for 
classical Skyrmions with baryon number $B=2,3, \cdots , 8, 17$ at $m_\pi=0$. }
\label{B>1chargedist}
\end{figure*}

In the following, we concentrate on the case with magnetic-field backgrounds $B_i$.   
The anomalous charge density is indeed induced in nucleons: 
\begin{align}
\langle j^0_{\rm anm}\rangle_{I_3,S_3}=\frac{ie^2N_c}{48\pi^2}B_i\langle P_i\rangle_{I_3,S_3}.
\end{align}
Figure~\ref{B1chargedist} shows the baryon number distribution, and the anomalous 
charge distribution under magnetic field 
along the 3rd- and the 1st-axes of quantized Skyrmions at $m_\pi=0$. 
The configurations of the charge distribution look like wave functions of an electron in a hydrogen atom. 

In contrast, we find that the matrix element of the spatial component of the current density vanishes, 
\begin{align}
\langle j_{{\rm anm}}^i \rangle_{I_3, S_3}
=0.\label{currentdist}
\end{align}
Thus, the electric current is not induced \cite{EHIIM2011}.

Let us calculate the total electric charge from the anomalous effect 
of $\langle P_i \rangle_{I_3,S_3}$ over the whole space 
gives
\begin{align}
\int d^3x \langle P_i \rangle_{I_3,S_3}&=
\left\{
\begin{array}{ll}
0 & (i=1,2), \\
-\dfrac{16\pi i}{9}(4I_3 S_3)c_0 & (i=3),
\end{array}
\right.\nonumber
\end{align}
where $c_0=\int dr \{r^2 f^\prime+\sin(2f)\}$. Its numerical value 
is $c_0=(-5.32, -12.3, -10.2, -7.32)$ for pion masses $m_\pi=(0, m_\pi^{\rm phys}/2, 
m_\pi^{\rm phys}, 2m_\pi^{\rm phys})$, respectively 
($m_\pi^{\rm phys} \equiv 0.263$ is the physical value of the pion mass in the unit of $eF_\pi$, determined from the mass splitting between nucleon and $\Delta$ \cite{Adkins1983}). 
We obtain the anomalous charge for nucleons  
\begin{align}
Q_{\rm anm}
=\frac{4e^2N_c}{27\pi}I_3 S_3\frac{c_0 B_3}{(e_sF_\pi)^2}.
\label{netcharge}
\end{align} 
Equation (\ref{netcharge}) shows that an electric charge is actually induced by the anomalous effect 
{\it even for a neutron}. 
We further find that dipole moment vanishes while 
quadrupole moment appears as a leading multipole \cite{EHIIM2011}.

\vspace{2mm}\noindent{\it No cancellation of the induced charge.} --- 
A large $N_c$ argument helps us to show that this anomaly-induced charge \eqref{netcharge} cannot be cancelled
by possible other electromagnetic corrections to the Skyrmion, as they are sub-leading in the 
$1/N_c$ expansion. In the Skyrme model (\ref{action}) itself, there are two possible electromagnetic 
corrections: (i) deformation of the Skyrmion configuration due to the magnetic field, and 
(ii) deformation of the Skyrme wave function via an induced potential in the quantum mechanics of 
the moduli fields $a_i$. 

First, we treat (i).
Since the Skyrme equations of motion is written by the normalized coordinate
$r/(e_s F_\pi)$ and $e_s F_\pi = {\cal O}(N_c^0)$, the classical deformation of the Skyrmion field 
$U$ should be ${\cal O}(e B N_c^0)$. So the effect of this classical deformation to the Gell-Mann-Nishijima
formula is 
\begin{align}
e I_3 \rightarrow e (I_3 + {\cal O}(eB N_c^0)), 
\label{cor}
\end{align}
which is smaller than the anomaly-induced
charge \eqref{netcharge} $Q_{\rm anm} = {\cal O}(e^2 B N_c^1)$ in the order of $N_c$. 
Second, from the action (\ref{action}), the induced potential in (ii) is   
the same order in $N_c$ as the kinetic term of the quantum mechanics. 
After including the correction, the $N_c$ order of the expectation value of $I_3$ 
remains intact, 
and then again Eq.~(\ref{cor}) follows. 

We conclude that
at large $N_c$ the anomaly-induced charge is at the leading order among the magnetic effects.
It is expected that even for finite $N_c$ an eventual cancellation does not occur.

\vspace{2mm}\noindent{\it Anomalous charges of multi-baryons (nucleus).} ---
We here calculate charge distributions of the 
classical Skyrmions with $B=2,3,\cdots 8, 17$ under magnetic fields. 
We consider classical Skyrmions because 
quantization of those with higher baryon number is generally difficult 
due to the complexity of their moduli space \footnote{
We can show a simple relation between classical and 
quantum anomalous current for $B=1$, that is, 
$\langle j^\mu_{{\rm anm}} \rangle_{I_3,S_3} =-\frac{4}{3}I_3 S_3 j^{\mu}_{{\rm anm}}[U_0],$
where $j^{\mu}_{{\rm anm}}[U_0]$ is the classical anomalous current for the Skyrme 
solution $U_0$ in Eq.~(\ref{genehedgehog}). 
This relation indicates that the classical result might not be meaningless. }.

To obtain the Skyrme solutions with higher baryon number $B\ge 2$, 
we use the standard rational-map ansatz \cite{Houghton1998}. 
The obtained solutions $f_B$ are substituted to the expression for 
classical charge distribution,
\begin{align}
j^0_{\rm anm}=\frac{ie^2N_c}{96\pi^2}\epsilon^{\mu\nu\rho\sigma}
F_{\nu\rho} P_\sigma[f_B],
\end{align}
where  $P_\sigma[f_B]$ is same as $P_\sigma$ in Eq.~(\ref{chargedistcls}) except that 
it consists of $f_B$ instead of $f$. 

Figure \ref{B>1chargedist} is the baryon number distribution, the charge distribution under magnetic field 
along the 3rd- and the 1st-axes, for baryon numbers $B=2,3, \cdots, 8, 17$. 
They show very amusing structures. 
The charge distribution of $B=2$ is like that of $B=1$, 
even with a difference in the density distribution of the baryon number. 
The charge distributions become more intricate as the baryon number $B$ increases. 
The baryon number density of $B=17$ has 
a structure of a fullerene C$_{60}$, and 
the charge distribution of (b) looks like a sea anemone 
and that of (c) looks like a pumpkin. 
It would be intriguing to see how these classical results are inherited to
observable charge distributions once the multi-Skyrmions are quantized.

\vspace{2mm}\noindent{\it Observation.} ---
Let us argue possibilities to observe the anomalous charge. 
First, we estimate the amount of induced charge in a nucleon. 
Using Eq.~(\ref{netcharge}), we obtain 
$Q_{\rm anm}\sim e\times 10^{-20}I_3 S_3{\rm [G^{-1}]}\times B_3 {\rm [G]}$. 
Under the terrestrial magnetic field $B\sim 1$[G], 
the induced charge $Q_{\rm anm}$ is about $10^{-20}e$, 
which would be too small to observe. 
On the surface of a magnetar, which is a neutron star with very strong magnetic field of order
$10^{15}$[G] \cite{Thompson1995}, $Q_{\rm anm}$ is about $10^{-5}e$. 
In heavy ion collisions, magnetic field of order $10^{17}$[G] 
 would be created \cite{Kharzeev2008}. 
However, $Q_{\rm anm}$ is about $10^{-3}e$ even for such an extremely strong magnetic field. 
Hence, it is natural that the electric charge of neutrons has never been detected until now. 

Next, electric dipole moment (EDM) of nucleons is not induced from the anomaly.  
This is consistent with the experimental results that there is no evidence for the existence of neutron EDM 
(see, e.g., \cite{Baker2006a}), which is performed under a magnetic field. 
In our study, the leading multipole is a quadrupole, 
$Q_{33}=-2Q_{11}=-2Q_{22}\sim e\times10^{-19} I_3 S_3 {\rm [fm^2 G^{-1}]} \times B_3 {\rm [G]}$. 
Its experimental measurement would be interesting. 

To see the universality of the generation of the anomalous charge, 
confirmation in other approaches is desirable. 
For instance, lattice QCD simulation with external electromagnetic fields is a reliable approach.
Holographic QCD is also helpful for gaining insights.

We have found that a neutron has a nonzero electric charge in external magnetic fields. 
Neutrons play an important role on the frontiers of hadron physics, such as neutron stars and heavy-ion collisions, 
where strong magnetic fields exist. 
Such neutrons would have anomalous charges which may be physically significant. 
Our results will bring new aspects of the dynamics of hadrons. 

\vspace*{2mm}
{\noindent \it Acknowledgment.} ---
The authors would like to thank Koichi Yazaki, Makoto Oka, and 
Nodoka Yamanaka for useful comments and discussions. 
The work of M.E.~is supported by Special Postdoctoral Researchers
Program at RIKEN. K.H.~and T. I.~are supported in part by the Japan 
Ministry of Education, Culture, Sports, Science and Technology.


\end{document}